\documentclass[sigconf]{acmart}

\usepackage{amsmath,amssymb,amsfonts}
\usepackage{algorithmic}
\usepackage{graphicx}
\usepackage{textcomp}
\usepackage{xcolor}
\usepackage{url}
\usepackage{booktabs}
\usepackage{multirow}
\usepackage{multicol}
\usepackage{tabularx}
\usepackage{listings}
\usepackage{pifont}
\usepackage{makecell}
\usepackage{ulem}
\usepackage{subcaption} 
\usepackage{dblfloatfix}
\usepackage{tcolorbox}
\usepackage[table]{xcolor}
\tcbuselibrary{skins, breakable} 

\PassOptionsToPackage{hyphens}{url} 
\AtBeginDocument{%
  }

\setcopyright{acmlicensed}
\copyrightyear{2018}
\acmYear{2025}
\acmDOI{XXXXXXX.XXXXXXX}
\acmConference[Conference acronym 'XX]{}{xxx xx--xx,
  xxxx}{xxx, xx}

\begin{document}


\title{Integrating LLM and Diffusion-Based Agents for Social Simulation}

\author{Xinyi Li}
\email{lixinyi22@mails.tsinghua.edu.cn}
\affiliation{
  \institution{DCST, Tsinghua University}
  \city{Beijing}
  \country{China}
}

\author{Zhiqiang Guo}
\email{georgeguo.gzq.cn@gmail.com}
\affiliation{
  \institution{DCST, Tsinghua University}
  \city{Beijing}
  \country{China}
}

\author{Qinglang Guo}
\email{gql1993@mail.ustc.edu.cn}
\affiliation{
  \institution{University of Science and Technology of China}
  \city{Hefei}
  \country{China}
}

\author{Hao Jin}
\email{jinhao1@cetc.com.cn}
\affiliation{
  \institution{China Academy of Electronics and Information Technology}
  \country{China}
}

\author{Weizhi Ma}
\email{mawz@tsinghua.edu.cn}
\affiliation{
  \institution{AIR, Tsinghua University}
  \city{Beijing}
  \country{China}
}

\author{Min Zhang}
\authornotemark[1]
\email{z-m@tsinghua.edu.cn}
\affiliation{
  \institution{DCST, Tsinghua University}
  \city{Beijing}
  \country{China}
}

\begin{abstract}
Large language models (LLMs) offer strong semantic reasoning capabilities for user modeling, but applying LLM-based simulation to an entire social network is computationally expensive and often unreliable for users with sparse behavioral histories. Meanwhile, conventional information diffusion models efficiently exploit historical propagation patterns and social structures, but provide limited understanding of item content and user-item semantic compatibility. We propose HySID, a hybrid framework for individual-level information adoption prediction that combines semantic reasoning with structural diffusion. HySID first analyzes the historical user-relation graph to adaptively select a small set of structurally informative core users. It then applies LLM-based simulation to estimate the engagement of these users and converts the judgments into a diffusion-compatible seed. Finally, a plug-in diffusion backbone propagates this seed through historical interaction structures to rank potential adopters across the full user population. This design enables LLMs to focus on users for whom semantic reasoning is most informative while allowing neural diffusion models to generalize the evidence to users that are not explicitly simulated. Experiments on three real-world datasets from Weibo, Zhihu, and KuaiRand show that HySID consistently improves four diffusion backbones in Recall and NDCG, while outperforming full-population LLM simulation baselines. At the same time, selective simulation reduces LLM inference cost by approximately 83\% to 99.9\%, demonstrating that HySID provides an effective and computationally practical approach to scalable information adoption prediction.
\end{abstract}




\maketitle

\section{Introduction}
Online social platforms have become primary channels for the spread of public opinion, news, and rumors. Information diffusion prediction~\cite{zhang_retweet_2015, yu_what_2011} is a research problem that models the propagation of information and predicts its effects on user engagement and public opinion. Given its direct relevance to social governance~\cite{rossettiSocialLLMpoweredSocial2024a}, rumor monitoring~\cite{zheng2023rumor}, and personalized recommendation~\cite{wuDiffNetNeuralInfluence2022}, information diffusion prediction has become increasingly important in the contemporary digital era.

The emergence of large language models (LLMs) has opened up new directions for this research field. LLMs can not only understand rich semantic information~\cite{zhang2024generative}, but also simulate human-like value judgments~\cite{lee2024enhancing}, emotional variations~\cite{liCAMELCommunicativeAgents2023}, and behavioral logic~\cite{wu2024deciphering}, thereby supporting the prediction of real-world social trends through user simulation. In recent years, some studies~\cite{gaoAgentScopeFlexibleRobust2024, bougieCitySimModelingUrban2025} have treated each agent as an autonomous individual and recreated real-world social dynamics by simulating the interactions between a number of agents under specific rules. 

However, applying LLM simulation to every user is both costly and unnecessary for information diffusion prediction. Diffusion scenarios usually involve large user populations and extensive interaction records, so assigning an LLM agent to every user incurs considerable computation and generation latency. More importantly, only users with rich historical information can benefit from the reasoning and general knowledge capabilities of LLMs. For the silent majority, the limited amount of behavioral evidence provides an insufficient basis for LLM reasoning, and the resulting predictions are therefore often less accurate. In addition, full-population LLM simulation typically relies on interactions among individual agents, which limits the joint modeling of implicit interaction dependencies and collaborative signals encoded in the social network.

These limitations motivate our hybrid design, which combines LLM-based simulation with a neural network-based diffusion model for accurate and efficient information diffusion prediction. Rather than asking LLM agents to assess every user, we apply semantic reasoning to a manageable subset and use the diffusion model to propagate the results through social links. This hybrid framework enables LLM-based simulation to concentrate its reasoning capacity within a limited inference budget, while the diffusion predictor extrapolates the predictions to the broader user population by leveraging implicit signals at a substantially lower cost. 

This complementary design leaves two central challenges. First, the users selected for LLM simulation must be identified adaptively. A fixed selection rule may fail because user relation structures and diffusion patterns vary across platforms and events, causing the same criterion to identify users with different levels of informativeness. The selection process should therefore exploit the structure of the user-relation graph to determine where limited reasoning is most valuable. Second, these two methods originally differed in input format and operating mode. LLM is designed to consume natural language inputs and produce decisions over user-item pairs. By contrast, traditional diffusion models operate on structured relational inputs and perform batch propagation. the semantic judgments of LLM-based simulation must be integrated with the neural network-based diffusion predictor in a principled way. Therefore, a principled integration framework that reconciles these representational and operational differences is required, enabling the two components to interact in a mutually reinforcement, rather than simply merging respective results together. 

To address these challenges, we propose HySID, an LLM-empowered hybrid simulation framework for information diffusion organized as a sequential three-agent pipeline. The Selection Agent uses a structure-driven selection skill to adaptively route among complementary user-selection strategies, determining where limited semantic reasoning should be allocated. The Simulation Agents apply LLM-based simulation to these users and organizes their item-specific interaction judgments into a pseudo-cascade, thereby producing a diffusion-compatible seed. The Diffusion Agent receives this seed through a standardized interface and passes it to a plug-in diffusion backbone, which uses social links and historical propagation patterns to extrapolate the simulation results to the remaining users and produce the final full-population ranking.

This decomposition assigns each type of reasoning to the stage where it is most useful. The Selection Agent adapts the allocation of limited semantic computation to the relational structure, the Simulation Agents contribute content-sensitive evidence that cannot be recovered from graph structure alone, and the Diffusion Agent exploits implicit collaborative signals to extend this evidence to the entire user population. The standardized seed interface also allows different diffusion backbones to be plugged into the final stage without changing the selection or simulation agents. Consequently, HySID combines the semantic understanding and world knowledge of LLMs with the scalability and collaborative modeling ability of diffusion prediction. The main contributions of this paper are summarized as follows.

\begin{enumerate}
    \item We introduce HySID, a sequential three-agent framework that organically bridges LLM-based user simulation and diffusion prediction models, jointly leveraging semantic behavioral reasoning and structural pattern modeling for information diffusion prediction.
    \item We propose a structure-driven core user selection procedure that adaptively routes among complementary user-selection strategies.
    \item Experiments on three real-world datasets show that HySID improves different diffusion backbones and outperforms LLM-based baselines at substantially lower cost than full-population simulation.
\end{enumerate}

\section{Related Work}
\subsection{LLM-based Social User Simulation}
The emergence of LLMs has spurred growing interest in social simulation, driven by the strong language understanding and generation capabilities of LLMs. Research in this area spans multiple scales~\cite{mou_individual_2024}, from simulations of single users to simulations of entire societies. At the single-user scale, many studies\cite{schwitzgebel2024creating, wang2024rolellm} have drawn analogies between the behavior of LLM agents and human behavioral patterns studied in psychology and neuroscience, seeking to align them in terms of correspondence, consistency, stability, and diversity. As the scope expands to multi-agent scenarios, researchers have investigated collaborative behaviors in goal-oriented settings, such as group debates~\cite{wang_rethinking_2024}, classroom interactions~\cite{chen2023agentverse}, and recommendation systems~\cite{wang_user_2024}. Finally, at the societal scale~\cite{piao_agentsociety_2025, wang_yulan-onesim_2025}, existing studies deploy large populations of LLM-based agents to act in turn and examine whether the resulting aggregate behavioral trends resemble real-world social phenomena.

However, LLM-driven social simulation faces two fundamental limitations that hinder its direct application to information diffusion prediction. The first is computational cost: a full-scale simulation in which each agent invokes an LLM for every information item would require millions of API calls for thousands of users and items, making such an approach economically prohibitive. The second limitation lies in the limited accuracy arising from the simulation mechanism. LLM agents may produce less reliable predictions for users with sparse behavioral records due to the agent need contextual information from different aspects for reasoning. This limitation is particularly important for the silent majority, whose limited behavioral records provide an insufficient basis for LLM. Consequently, large-scale LLM simulation may perform poorly on information diffusion prediction, which requires accurate predictions for specific users rather than only realistic population-level trends.


\subsection{Traditional Information Diffusion Prediction}
Modeling information diffusion processes in social networks is widely used in various real-world applications, including popularity prediction\cite{cheng2024retrieval}, micro-behaviour prediction\cite{wang2024information} and macro-size prediction\cite{xu2021casflow}. Early methods such as IC~\cite{kempe2003maximizing} and LT~\cite{granovetter1978threshold} rely on predefined stochastic propagation rules to evaluate the transfer probability between pairs of users. Other studies have carefully designed handcrafted cascade features that calculate diffusion probabilities using well-defined functions. With the development of deep learning techniques, existing methods learn user dependencies and implicit diffusion patterns from cascades to predict which users will spread the information. Some approaches model cascades as temporal sequences and captures user dependencies through RNN\cite{sherstinsky2020fundamentals,yang2018neural}, LSTM\cite{yu2019review, wang2017topological}, or attention mechanisms\cite{vaswani2017attention, islam2018deepdiffuse}. Other type models cascades as graphs or hypergraphs to capture the complex relational structures, using architectures such as GCN\cite{yuan2020dyhgcn} and HGAT\cite{sun2022ms}. Besides, some method incorporates auxiliary information, such as static social networks based on friendship or following relationships, alongside the diffusion graph\cite{yang_multi-scale_2019}. 

However, these neural network models rely primarily on historical interaction signals and provide limited modeling of information content. Their relatively inflexible integration of interaction and content signals also restricts their ability to predict user behavior based on the alignment between user interests and information content. In addition, they generally have weaker language understanding and reasoning capabilities than LLMs, which limits their ability to interpret content and infer user responses, especially for newly released items with limited diffusion histories.

\section{Methodology}

\subsection{Problem Definition}
Let $\mathcal{U} = \{ u_1, u_2, \dots, u_M \}$ denote the set of users in a social network. Each information item is an element $p \in \mathcal{P}$ and $\mathcal{P}$ is the collection of all information items. Every $p$ is associated with a tuple $(\text{content}, \text{publisher})$, where $\text{content}$ denotes the textual content of the item and $\text{publisher}$ denotes the user who originally posted it.

Besides, user history is also provided. For each user $u_i$, let $h_i = \{ (p_{i_{1}}, t_{i_{1}}), (p_{i_{2}}, t_{i_{2}}), \dots, (p_{i_{N_i}}, t_{i_{N_i}}) \}$ denote the complete interaction history of $u_i$, where each pair consists of an information item $p_{i_j} \in \mathcal{P}$ that $u_i$ interacted with and the corresponding timestamp $t_{i_j}$. The collective history of all users is denoted $\mathcal{H} = \{ h_1, h_2, \dots, h_M \}$.

Given the set $\mathcal{P}$ of information items and the complete histories $\mathcal{H}$ for all users in $\mathcal{U}$, we study new-item adopter ranking: for each $p_k \in \mathcal{P}$, we predict the subset of users in $\mathcal{U}$ who become activated and interact with $p_k$.

\subsection{Overview of the HySID Framework}

\begin{figure*}[htbp]
  \centering
  \includegraphics[width=0.92\textwidth]{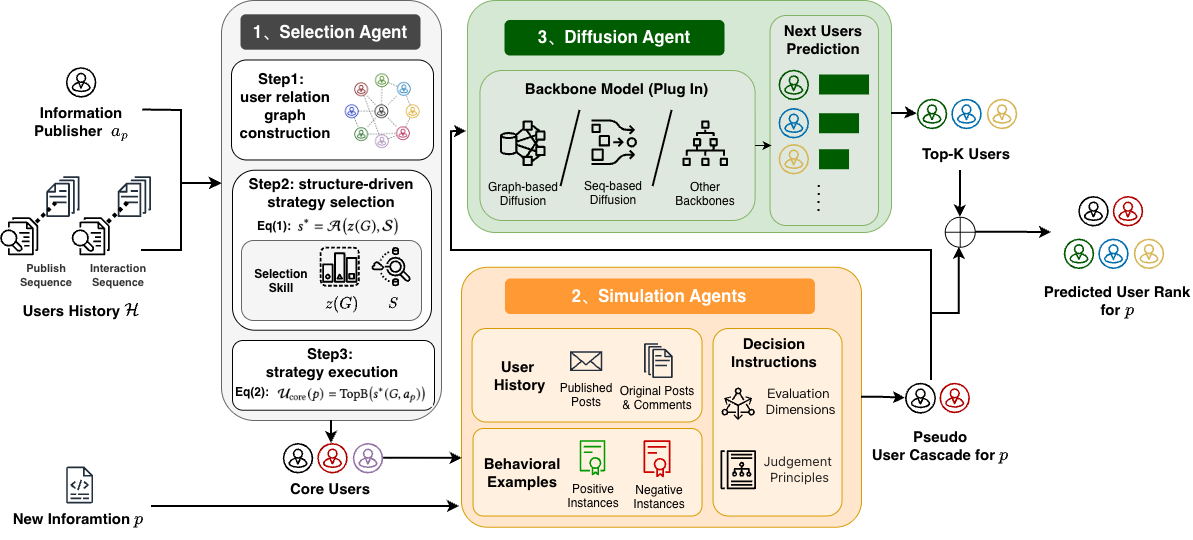}
  \caption{Overview of the proposed HySID framework, consisting of three types of agents: (i) the Selection Agent; (ii) the Simulation Agents; and (iii) the Diffusion Agent.}
  \label{fig:framework}
\end{figure*}

As illustrated in Figure~\ref{fig:framework}, HySID is a three-stage pipeline with coordinated agents for structure-driven core-user selection, LLM-based core-user interaction simulation, and full-population information diffusion prediction. 

In the first stage, the Selection Agent identifies a core-user set $\mathcal{U}_{\mathrm{core}}(p)$ whose adoption in later stages can be more effectively promoted through structural diffusion patterns. Considering that the underlying user relations differ substantially across platforms, the Selection Agent uses a structure-conditioned selection skill to adaptively determine the suitable strategy from graph-level diagnostic statistics computed on the user relation graph. This design allows the selected core users to better reflect the structural characteristics of diffusion on each dataset.

In the second stage, the Simulation Agents evaluate user-content alignment to determine whether the target information item is likely to trigger engagement, and produces a pseudo-cascade $C_{\mathrm{seed}}(p)$. The Simulation Agents draw on multiple sources of evidence, including the user's historical interactions, few-shot behavioral examples constructed offline, and multi-factor evaluation of the user's potential alignment with the target item. The resulting $C_{\mathrm{seed}}(p)$ serves as an intermediate representation that connects semantic user simulation with downstream structural propagation. The prompts for these two stages are in the anonymized code\footnote{\url{https://anonymous.4open.science/r/HySID-4AC6}}.

In the third stage, the Diffusion Agent conditions a plug-in diffusion backbone on $C_{\mathrm{seed}}(p)$ and predicts the behavior of the full user population by modeling structural and sequential diffusion patterns. The final cascade combines the seed users with the top-$K$ users ranked by the diffusion predictor. Thus, HySID applies semantic reasoning to structurally informative users and efficiently extends the prediction to the remaining population.

\subsection{Structure-driven Core User Selection}
\label{sec:structure-driven-selection}

The Selection Agent adaptively determines how to allocate the limited LLM inference budget across users according to the structure of the historical user relation graph. A fixed user-selection strategy is restrictive because platforms may differ a lot. HySID therefore performs structure-driven core-user selection through a unified selection skill, allowing the agent to use LLM's reasoning and knowledge abilities. The skill integrates three steps to achieve the adaptive user selection: it summarizes the historical user relation graph into a structural feature profile, applies the structure-driven policy to select a compatible user-prioritization strategy, and executes the selected strategy to construct the core-user set. 

To be specific, let $G$ be the directed weighted user relation graph constructed from historical interactions, and let $\mathcal{S}=\{s_1,\ldots,s_L\}$ denote a family of core-user selection strategies. Given the structural feature profile $z(G)$, the skill applies the structure-driven policy to select one strategy:
\begin{equation}
s^* = \mathcal{A}\bigl(z(G),\mathcal{S}\bigr)
\end{equation}
where $\mathcal{A}$ denotes the adaptive policy and $s^*$ is the selected strategy. 

We construct the user relation graph $G$ from user history $\mathcal{H}$, where nodes denote users and edges encode interactions. The graph is constructed based on the training data to avoid information leakage, and is used as the structural basis for feature extraction and core-user selection. For each cascade, the first user is treated as the diffusion root. We add a directed edge from the root user to each subsequently activated user, and further connect users who co-occur within a fixed temporal window. Edge weights accumulate across cascades. In this way, $G$ captures both root-centered propagation dependency and local co-participation patterns. 

The feature profile $z(G)$ and the policy used in the skill are designed to characterize the structural conditions under which different user-prioritization strategies may be informative. We predefine four complementary strategies, spanning global versus publisher-conditioned, and simple versus richer structural, notions of early semantic value. Tables~\ref{tab:selection-policy} and~\ref{tab:selection-strategies} summarize the policy categories, structural features, and strategies used in the selection skill. Other details are provided in the appendix ~\ref{appendix_skill} and the anonymized code.
\begin{table}[t]
  \centering
  \small
  \setlength{\tabcolsep}{3pt}
  \renewcommand{\arraystretch}{1.12}
  \caption{Policy categories and structural features used by the Selection Agent.}
  \label{tab:selection-policy}
  \begin{tabularx}{\columnwidth}{@{}>{\raggedright\arraybackslash}p{0.38\columnwidth} >{\raggedright\arraybackslash}X@{}}
    \toprule
    \textbf{Policy category} & \textbf{Structural feature} \\
    \midrule
    \multirow{4}{=}{\textbf{Evidence scope}\\[-2pt]\scriptsize Global vs. publisher-centered}
      & Average user activity \\
      & Mean weighted connectivity \\
      & Publisher path coverage \\
      & Median publisher-centered reachability \\
    \midrule
    \multirow{3}{=}{\textbf{Structural reliability}\\[-2pt]\scriptsize Connectivity coverage and concentration}
      & Activity concentration \\
      & Weighted connectivity concentration \\
      & Isolated-user ratio \\
    \midrule
    \multirow{2}{=}{\textbf{Ranking discrimination}\\[-2pt]\scriptsize Direct hop vs. richer multi-hop}
      & Shortest-path tie ratio \\
      & Within-hop PPR dispersion \\
    \bottomrule
  \end{tabularx}
\end{table}

\begin{table}[t]
  \centering
  \small
  \setlength{\tabcolsep}{3pt}
  \renewcommand{\arraystretch}{1.12}
  \caption{Deterministic strategies available to the Structure-driven selection skill.}
  \label{tab:selection-strategies}
  \begin{tabularx}{\columnwidth}{@{}>{\raggedright\arraybackslash}p{0.31\columnwidth} >{\raggedright\arraybackslash}X@{}}
    \toprule
    \textbf{Strategy} & \textbf{Ranking principle and structural rationale} \\
    \midrule
    \textbf{Global Activity}
      & Rank users by historical interaction activity; useful when publisher-conditioned paths are sparse or fragmented. \\
    \textbf{Global Centrality}
      & Rank users by weighted in-degree and out-degree; useful when weighted connectivity forms a stable global structure. \\
    \textbf{Publisher-Hop}
      & Rank users by directed shortest-path distance from the publisher; useful when direct hop distance separates candidates. \\
    \textbf{Publisher-Proximity}
      & Rank users by weighted personalized PageRank proximity from the publisher; useful when within-hop proximity provides additional differentiation. \\
    \bottomrule
  \end{tabularx}
\end{table}

The policy does not select users directly; instead, it determines which prioritization principle should be applied by the skill. Rather than associating one feature with one fixed decision, the policy considers complementary feature groups, therefore the feature--strategy correspondence specifies the structural assumptions of the available strategies rather than a manually enumerated decision tree: publisher coverage and reachability indicate whether publisher-centered strategies are informative, while hop ties and within-hop proximity distinguish direct path ranking from richer weighted proximity. Moreover, the framework is extensible: the Selection Agent's strategy set and policy can be adapted to different settings and requirements by registering alternative deterministic ranking functions in the skill, without changing the roles of the Simulation Agents and the Diffusion Agent.

Given the selected strategy and an externally fixed LLM inference budget $B$ (usually a threshold for the core-user set size), the skill then constructs a ranking over candidate users and returns the core-user set for information item $p$:
\begin{equation}
\mathcal{U}_{\mathrm{core}}(p)
 = \operatorname{TopB}\bigl(s^*(G,a_p)\bigr)
\end{equation}

where $a_p$ denotes the publisher of item $p$. For global strategies, the final ranking is independent of $a_p$, while for publisher-conditioned strategies, $a_p$ determines the information-specific ranking. The selected core users are then passed to the Simulation Agents for interaction judgment.

\subsection{LLM-based Core User Interaction Simulation}
After core user selection, the Simulation Agents conduct LLM-based user interaction simulation to estimate whether selected users are likely to engage with the target information item. For each $u\in\mathcal{U}_{\mathrm{core}}(p)$, the agent receives the target information content $\text{content}_p$, the user's history $h_u$, and a small set of offline-constructed behavioral examples through a structured prompt. It then returns an engagement score and a binary interaction decision $\widehat{y}_{p,u}$.

The simulation decision represents likely engagement inferred from available text and interaction history, combining multiple sources of user-specific behavioral evidence as follows:

\paragraph{The user history.} The user history contains previously published and interacted-with information items, together with the available interaction content, allowing the agent to infer observable patterns of topical interest, stylistic preference, and engagement tendency. When a user has insufficient historical interactions to support a reliable behavioral characterization, the Simulation Agents treat the user as inactive by default rather than making a confident semantic judgment from sparse evidence. 

\paragraph{The behavioral examples.} The behavioral examples include both positive cases drawn from the user's historical interactions and negative cases that were not adopted by the user. These examples provide user-specific behavioral references and help the agent distinguish engagement signals from superficial semantic similarity. The construction of these examples is performed offline, and details are provided in the appendix ~\ref{appendix_examples} and the anonymized code.

All historical records and behavioral examples used by the Simulation Agents are constructed exclusively from the predefined training and validation splits to avoid information leakage. 

The Simulation Agents evaluate the compatibility between the user and the target item from multiple complementary perspectives, including topical alignment, emotional resonance, stylistic compatibility, and potential inhibiting factors such as redundancy, ambiguity, or conflicts with the user's inferred preferences. The resulting score reflects the joint evidence across these dimensions rather than any single signal. When different dimensions provide conflicting evidence, the agent reduces its confidence to reflect the unresolved uncertainty. The resulting decision is therefore treated as an observable proxy for likely engagement inferred from available content and interaction history, rather than as direct access to the user's latent psychological state. The implemented prompt details are provided in the anonymized code.

The users identified as likely to engage are combined with the publisher to form the pseudo-cascade passed to the Diffusion Agent. The Simulation Agents output an interaction score, a binary decision, and a supporting rationale. A core user is included in the pseudo-cascade only when the predicted decision is positive and the interaction score exceeds the predefined threshold. Specifically, we define the seed state as
\begin{equation}
C_{\mathrm{seed}}(p)
=
\{a_p\}
\cup
\left\{
u\in\mathcal{U}_{\mathrm{core}}(p):
d_{p,u}=\mathrm{yes}
\land
s_{p,u}\geq\tau
\right\}
\end{equation}
By querying only the selected core users, the Simulation Agents supply content-sensitive evidence for the early diffusion state while keeping the number of LLM-based judgments bounded by the selection budget. The resulting $C_{\mathrm{seed}}(p)$ serves as the intermediate representation connecting user-level semantic simulation with downstream population-level diffusion prediction.

\subsection{Full User Information Diffusion Prediction}

The Diffusion Agent receives the pseudo-cascade generated by the Simulation Agents and extends it to the full user population. This design assigns semantically intensive early-stage judgments to the LLM module, while delegating large-scale structural generalization to a diffusion backbone.

Specifically, the publisher $a_p$ and $C_{\mathrm{seed}}(p)$ are organized into an early cascade, which is then provided to a plug-in diffusion backbone to score users who are not already in the seed state. Since the pseudo-cascade has the same format as the historically observed cascades, any diffusion model can be directly used as the backbone without changing the surrounding framework. Let $D_{\omega}$ denote a diffusion backbone and $R_{\mathrm{rest}}(p)$ its ranking over unseeded users:
\begin{equation}
 R_{\mathrm{rest}}(p)=D_{\omega}\bigl(G,C_{\mathrm{seed}}(p)\bigr).
\end{equation}
The final ranking $R_p$ is constructed by first ordering the core users predicted by the Simulation Agents according to their LLM-generated scores, appending the remaining users ranked by the Diffusion Agent, and retaining the top-$K$ users from the resulting list. Specifically, the final ranking is defined as
\begin{equation}
R_p
=
\operatorname{TopK}
\left(
\operatorname{Sort}_{\mathrm{score}}
\bigl(R_{\mathrm{LLM}}(p)\bigr)
\mathbin{\Vert}
R_{\mathrm{rest}}(p)
\right)
\end{equation}

where $\mathbin{\Vert}$ denotes ordered concatenation, and users included in
$R_{\mathrm{LLM}}(p)$ are masked from $R_{\mathrm{rest}}(p)$ to avoid
duplicate predictions. This interface leaves the internal architecture of $D_{\omega}$ unchanged and allows the diffusion backbone to exploit social links and historical propagation patterns to complete the individual-level prediction over the entire population.

\section{Experiments}
\subsection{Experimental Setup}
\subsubsection{Datasets}
\begin{table}[t]
  \centering
  \caption{Statistic information about dataset.}
  \label{tab:dataset-info}
  \begin{tabular}{lccccc}
    \toprule
    Dataset & \#Info.Threads & \#Users & \#Inter. & Avg. Len. \\
    \midrule
    Weibo & 273 & 6,247 & 7,550 & 27.66  \\
    Zhihu & 3,327 & 30,325 & 41,007 & 12.33  \\
    KuaiRand & 2,576 & 25,213 & 252,494 & 98.02 \\
    \bottomrule
  \end{tabular}
\end{table}
We conduct experiments on three real-world datasets from different social platforms: \textbf{Weibo}, \textbf{Zhihu}~\cite{hao2021largescale}, and \textbf{KuaiRand}~\cite{gao2022kuairand}. The Zhihu and KuaiRand dataset are public available while the Weibo dataset is constructed by ourselves from several recent public discussion topics on Sina Weibo platform. For KuaiRand, where the original short-video records lack textual descriptions, we construct information content from video titles, categories, and cover text, and treat user--video interactions as activations. Dataset statistics are summarized in Table~\ref{tab:dataset-info}. Across all datasets, repeated interactions with the same item are reduced to the earliest interaction. We also remove cascades with fewer than three interactions to reduce the influence of highly incidental; this cascade-level filter is distinct from the limited user-history setting caused by newly published information items, which concerns the amount of historical information available for a user behaviour prediction.


\subsubsection{Baseline Methods}
We compare our framework against two categories of baseline methods: (1) \textit{general LLM-based social simulators}, which directly model user behaviors through LLM-driven agents without task-specific diffusion modeling; and (2) \textit{traditional information diffusion prediction models}, which learn structural and temporal diffusion patterns from historical cascades. The diffusion prediction models are evaluated both as standalone baselines and as plug-in backbones within our hybrid framework to assess the generality of the proposed design.

The LLM-based general social simulators include:
\begin{enumerate}
    \item \textbf{AgentSociety}~\cite{piao_agentsociety_2025}: A large-scale LLM-driven social simulation framework that models human-like agents in virtual environments, emphasizing emergent social behaviors through large-scale generative simulation.

    \item \textbf{YuLan-OneSim}~\cite{wang2025yulan}: A general LLM-based social simulator that supports configurable simulation through natural language descriptions and scalable multi-agent interactions. 
\end{enumerate}

The traditional information diffusion prediction models include:

\begin{enumerate}
    \item \textbf{DyHGCN}~\cite{yuan2020dyhgcn}: A dynamic heterogeneous graph convolutional network that models temporal diffusion graphs by encoding dynamic dependencies into user representations.

    \item \textbf{DisenIDP}~\cite{cheng_enhancing_2023}: A disentangled diffusion prediction model that captures latent user intents from both interest-wise and dependency-wise perspectives, while modeling both long-term and short-term cascade influences.

    \item \textbf{RotDiff}~\cite{qiao2023rotdiff}: A diffusion model that employs rotation transformations in hyperbolic space to capture complex structural diffusion patterns.

    \item \textbf{MIM}~\cite{cheng2025disentangling}: A multi-channel hypergraph-based diffusion model that incorporates hierarchical knowledge distillation to capture both historical and future diffusion signals.
\end{enumerate}

\subsubsection{Evaluation Metrics}
We formulate information diffusion prediction as a top-$k$ adopter-ranking task and report \textbf{Recall@}k and \textbf{NDCG@}k. Recall measures the coverage of activated users, while NDCG additionally considers their ranking positions. We evaluate both metrics at $k\in\{50,100\}$ using the same cutoffs for all methods.

\subsubsection{Implementation Details}
Following the common practice in information diffusion prediction, we partition each dataset into training, validation, and test splits with a ratio of 8:1:1. Diffusion models are trained exclusively on the training split, whereas the validation split is used for model selection and early stopping. For the LLM-based components, user histories and behavioral examples are constructed from information available in the training and validation splits, without using future interactions from the test split. All methods are evaluated on the same test items and candidate user population. For each target item, the available item-level inputs are the publisher ID, publication timestamp, and textual content. We apply the same top-$k$ prediction budget to all methods. Since AgentSociety and YuLan-OneSim produce binary decisions rather than ranking scores, we retain their native output order and use the first $k$ positively predicted users when more than $k$ positive users are produced.

For all LLM-based methods, we use Qwen3-8B with an 8192-token context limit, temperature 0.8, and top-$p$ 0.95, and deploy inference locally using vLLM. For LLM-based simulation, we retain at most 40 historical interaction records per user.

For HySID, the Selection Agent's feature profile $z(G)$ are computed from the train and validation split. The number of selected core users is set to a fixed budget equal to $1\%$ of the total user population for each dataset. Users with fewer than three available historical interactions are treated as inactive by default. For the remaining users, we use at most three positive and three negative behavioral examples and set the interaction-score threshold to 5 on a 10-point scale. The examples and their supporting rationales are annotated offline using GPT-4o-mini with information from the training and validation splits only.

For AgentSociety, each agent is initialized with historical user behaviors as memory, and simulated interaction decisions are collected through its built-in survey mechanism. For YuLan-OneSim, we use its built-in \textit{information cascade and silence} scenario and initialize users with an \texttt{interaction\_log} constructed from historical publishing and interaction behaviors. Owing to its higher runtime overhead, YuLan-OneSim is evaluated only on Weibo.

For diffusion-based methods, we use cross-entropy loss and train each model for at most 200 epochs, with early stopping when the validation performance does not improve for 10 consecutive epochs. We use Adam with a batch size of 64, a learning rate of $10^{-3}$, and an embedding dimension of 64; other backbone-specific settings follow the corresponding original implementations.

All experiments are conducted on an NVIDIA A800-SXM4-80GB GPU server. The anonymized code is available\footnote{\url{https://anonymous.4open.science/r/HySID-4AC6}}.

\subsection{Overall Performance}

\begin{table*}[t]
  \centering
  \caption{Overall performance and per-item computation cost. All methods are evaluated on the same test set with the same top-$k$ cutoffs. $^{**}$ and $^{*}$ indicate $p<0.01$ and $p<0.05$, respectively, under a paired permutation test against the corresponding diffusion backbone. Boldface compares each HySID+ variant with its paired backbone baseline. Costs are averaged per information item, the lower the better. Parenthesized reductions for HySID+ are calculated relative to the second-best cost baseline result. The merged diffusion columns report the LLM cost of HySID. $^{\dagger}$ denotes an extrapolated estimate based on sampled result; ``--'' denotes rows for which $k$ is unavailable.}
  \label{tab:overall-wsdm}
  \small
  \setlength{\tabcolsep}{2.6pt}
  \resizebox{\linewidth}{!}{%
  \begin{tabular}{c|cc|cc|cc|cc|cc|cc}
    \toprule
    \multirow{2}{*}{Dataset} & \multirow{2}{*}{Metric} & \multirow{2}{*}{$k$}
      & \multicolumn{2}{c|}{LLM-based}
      & \multicolumn{2}{c|}{DyHGCN}
      & \multicolumn{2}{c|}{DisenIDP}
      & \multicolumn{2}{c|}{RotDiff}
      & \multicolumn{2}{c}{MIM} \\
    \cmidrule(lr){4-5}\cmidrule(lr){6-7}\cmidrule(lr){8-9}\cmidrule(lr){10-11}\cmidrule(lr){12-13}
      & & & AgentSociety & YuLan-OneSim
      & Baseline & \cellcolor{gray!15}HySID+
      & Baseline & \cellcolor{gray!15}HySID+
      & Baseline & \cellcolor{gray!15}HySID+
      & Baseline & \cellcolor{gray!15}HySID+ \\
    \midrule

    \multirow{7}{*}{Weibo}
      & \multirow{2}{*}{Recall $\uparrow$} & 50 & 0.0029 & 0.0219 & 0.1190 & \cellcolor{gray!15}\textbf{0.1773}$^{*}$ & 0.0419 & \cellcolor{gray!15}\textbf{0.1772}$^{**}$ & 0.0254 & \cellcolor{gray!15}\textbf{0.1793}$^{**}$ & 0.0740 & \cellcolor{gray!15}\textbf{0.1813}$^{**}$ \\
      & & 100 & 0.0051 & 0.0303 & 0.2145 & \cellcolor{gray!15}\textbf{0.2684}$^{*}$ & 0.0635 & \cellcolor{gray!15}\textbf{0.2539}$^{**}$ & 0.0447 & \cellcolor{gray!15}\textbf{0.2703}$^{**}$ & 0.1412 & \cellcolor{gray!15}\textbf{0.2684}$^{**}$ \\
    \cmidrule(lr){2-13}
      & \multirow{2}{*}{NDCG $\uparrow$} & 50 & 0.0025 & 0.0109 & 0.0927 & \cellcolor{gray!15}\textbf{0.1852}$^{**}$ & 0.0306 & \cellcolor{gray!15}\textbf{0.1849}$^{**}$ & 0.0180 & \cellcolor{gray!15}\textbf{0.1862}$^{**}$ & 0.0578 & \cellcolor{gray!15}\textbf{0.1871}$^{**}$ \\
      & & 100 & 0.0036 & 0.0140 & 0.1296 & \cellcolor{gray!15}\textbf{0.2290}$^{**}$ & 0.0407 & \cellcolor{gray!15}\textbf{0.2218}$^{**}$ & 0.0289 & \cellcolor{gray!15}\textbf{0.2297}$^{**}$ & 0.0926 & \cellcolor{gray!15}\textbf{0.2290}$^{**}$ \\
    \cmidrule(lr){2-13}
      & \multirow{3}{*}{Cost $\downarrow$} & $N_{\mathrm{call}}$ & 12,494 & 50,132 & \multicolumn{8}{c}{\cellcolor{gray!15} $N_{\mathrm{call}}$ of HySID+: \textbf{63} (99.50\% reduction)} \\
      & & Tokens & 6.96M & 64.72M & \multicolumn{8}{c}{\cellcolor{gray!15} Tokens of HySID+: \textbf{0.19M} (97.27\% reduction)} \\
      & & Time (s) & 234.4 & 23,872.4 & \multicolumn{8}{c}{\cellcolor{gray!15} Time of HySID+: \textbf{6.5} (97.23\% reduction)} \\
    \midrule

    \multirow{7}{*}{Zhihu}
      & \multirow{2}{*}{Recall $\uparrow$} & 50 & 0.0025 & {--} & 0.0235 & \cellcolor{gray!15}\textbf{0.0370}$^{**}$ & 0.0022 & \cellcolor{gray!15}\textbf{0.0280}$^{**}$ & 0.0193 & \cellcolor{gray!15}\textbf{0.0353}$^{**}$ & 0.0275 & \cellcolor{gray!15}\textbf{0.0356}$^{**}$ \\
      & & 100 & 0.0025 & {--} & 0.0359 & \cellcolor{gray!15}\textbf{0.0472}$^{*}$ & 0.0040 & \cellcolor{gray!15}\textbf{0.0350}$^{**}$ & 0.0229 & \cellcolor{gray!15}\textbf{0.0468} & 0.0381 & \cellcolor{gray!15}\textbf{0.0481}$^{**}$ \\
    \cmidrule(lr){2-13}
      & \multirow{2}{*}{NDCG $\uparrow$} & 50 & 0.0017 & {--} & 0.0122 & \cellcolor{gray!15}\textbf{0.0241}$^{**}$ & 0.0008 & \cellcolor{gray!15}\textbf{0.0200}$^{**}$ & 0.0087 & \cellcolor{gray!15}\textbf{0.0236}$^{**}$ & 0.0138 & \cellcolor{gray!15}\textbf{0.0233}$^{**}$ \\
      & & 100 & 0.0016 & {--} & 0.0149 & \cellcolor{gray!15}\textbf{0.0270}$^{**}$ & 0.0014 & \cellcolor{gray!15}\textbf{0.0218}$^{**}$ & 0.0089 & \cellcolor{gray!15}\textbf{0.0272}$^{**}$ & 0.0157 & \cellcolor{gray!15}\textbf{0.0270}$^{**}$ \\
    \cmidrule(lr){2-13}
      & \multirow{3}{*}{Cost $\downarrow$} & $N_{\mathrm{call}}$ & 60,650 & 241,761$^{\dagger}$ & \multicolumn{8}{c}{\cellcolor{gray!15} $N_{\mathrm{call}}$ of HySID+: \textbf{1450} (97.61\% reduction)} \\
      & & Tokens & 43.12M & 301.23M$^{\dagger}$ & \multicolumn{8}{c}{\cellcolor{gray!15} Tokens of HySID+: \textbf{5.46M} (87.34\% reduction)} \\
      & & Time (s) & 2,580.9 & 115,124.3$^{\dagger}$ & \multicolumn{8}{c}{\cellcolor{gray!15} Time of HySID+: \textbf{437} (83.07\% reduction)} \\
    \midrule

    \multirow{7}{*}{KuaiRand}
      & \multirow{2}{*}{Recall $\uparrow$} & 50 & {0.0024} & {--} & 0.0251 & \cellcolor{gray!15}\textbf{0.0360}$^{*}$ & 0.0433 & \cellcolor{gray!15}\textbf{0.0454} & 0.0378 & \cellcolor{gray!15}\textbf{0.0484}$^{*}$ & 0.0328 & \cellcolor{gray!15}\textbf{0.0443}$^{*}$ \\
      & & 100 & {0.0030} & {--} & 0.0410 & \cellcolor{gray!15}\textbf{0.0579}$^{**}$ & 0.0637 & \cellcolor{gray!15}\textbf{0.0813}$^{**}$ & 0.0623 & \cellcolor{gray!15}\textbf{0.0791}$^{**}$ & 0.0471 & \cellcolor{gray!15}\textbf{0.0654}$^{**}$ \\
    \cmidrule(lr){2-13}
      & \multirow{2}{*}{NDCG $\uparrow$} & 50 & {0.0050} & {--} & 0.0264 & \cellcolor{gray!15}\textbf{0.0342} & 0.0397 & \cellcolor{gray!15}\textbf{0.0418} & 0.0335 & \cellcolor{gray!15}\textbf{0.0411}$^{*}$ & 0.0312 & \cellcolor{gray!15}\textbf{0.0372} \\
      & & 100 & {0.0044} & {--} & 0.0297 & \cellcolor{gray!15}\textbf{0.0409}$^{*}$ & 0.0420 & \cellcolor{gray!15}\textbf{0.0534} & 0.0408 & \cellcolor{gray!15}\textbf{0.0511}$^{**}$ & 0.0344 & \cellcolor{gray!15}\textbf{0.0440}$^{*}$ \\
    \cmidrule(lr){2-13}
      & \multirow{3}{*}{Cost $\downarrow$} & $N_{\mathrm{call}}$ & 50,426 & 204,431$^{\dagger}$ & \multicolumn{8}{c}{\cellcolor{gray!15} $N_{\mathrm{call}}$ of HySID+: \textbf{17} (99.97\% reduction)} \\
      & & Tokens & 39.28M & 279.46M$^{\dagger}$ & \multicolumn{8}{c}{\cellcolor{gray!15} Tokens of HySID+: \textbf{0.06M} (99.85\% reduction)} \\
      & & Time (s) & 2,211.7 & 102,215.5$^{\dagger}$ & \multicolumn{8}{c}{\cellcolor{gray!15} Time of HySID+: \textbf{6.3} (99.72\% reduction)} \\
    \bottomrule
  \end{tabular}}
\end{table*}

Table~\ref{tab:overall-wsdm} reports the overall performance of all methods. Because the LLM-based baselines produce binary predictions without ranking, their NDCG values are computed using the default output order. These values are therefore reported only as an auxiliary reference. Due to the high cost of YuLan-OneSim, we do not execute YuLan-OneSim on Zhihu and KuaiRand and only report its estimated cost based on small-sample measurements.

The results show that HySID+ consistently improves every diffusion backbone across all datasets and metrics, regardless of the standalone strength of the backbone. These gains demonstrate the effectiveness and broad applicability of HySID. Moreover, the improvements in NDCG indicate that HySID not only retrieves more activated users but also ranks relevant users higher. These results suggest that HySID leverages the complementary predictive strengths of the Simulation Agents and the Diffusion Agent. By using structural diffusion to extend simulation result, the Simulation Agents produce reliable predictions for selected users. Meanwhile, the simulated users provide a semantic seed cascade that gives diffusion prediction a richer starting state than the sparse early cascade of a newly published item alone, therefore achieving better performance than either component used alone.

Table~\ref{tab:overall-wsdm} also compares the computational cost of different methods where all metrics are reported per post. To ensure a comparable inference environment, AgentSociety and HySID were executed with the same maximum number of LLM requests, set to 300. YuLan-OneSim does not expose a unified maximum-concurrency control for LLM requests in its default implementation, which are scheduled according to the simulator's native multi-agent execution mechanism. We therefore retained YuLan-OneSim's default execution behavior rather than introducing an external concurrency constraint. All methods were evaluated on the same hardware, and the reported wall-clock time reflects their actual end-to-end execution under these respective native settings.

In particular, both AgentSociety and YuLan-OneSim require LLM simulation over full user population, resulting in very large numbers of requests and token consumption. The efficiency of HySID comes from its Selection Agent, which reduces the number of LLM calls by two to three orders of magnitude while preserving full-population prediction. More importantly, this reduction does not sacrifice full-population prediction. The additional cost of this diffusion stage is negligible in practice, since diffusion inference requires only about 1-4 second per epoch on average. It is worth noting that the selection strategy is dataset-specific, so the number of simulation times is not a fixed proportion of the total population. This explains why KuaiRand and Zhihu incur substantially different LLM costs despite having comparable user populations. Overall, HySID provides more effective and computationally practical full-population prediction than direct LLM simulation of all users.

\subsection{Ablation Study}
\subsubsection{User Selection Strategy Comparison}
\begin{figure}[t]
  \centering
  \includegraphics[width=0.45\textwidth]{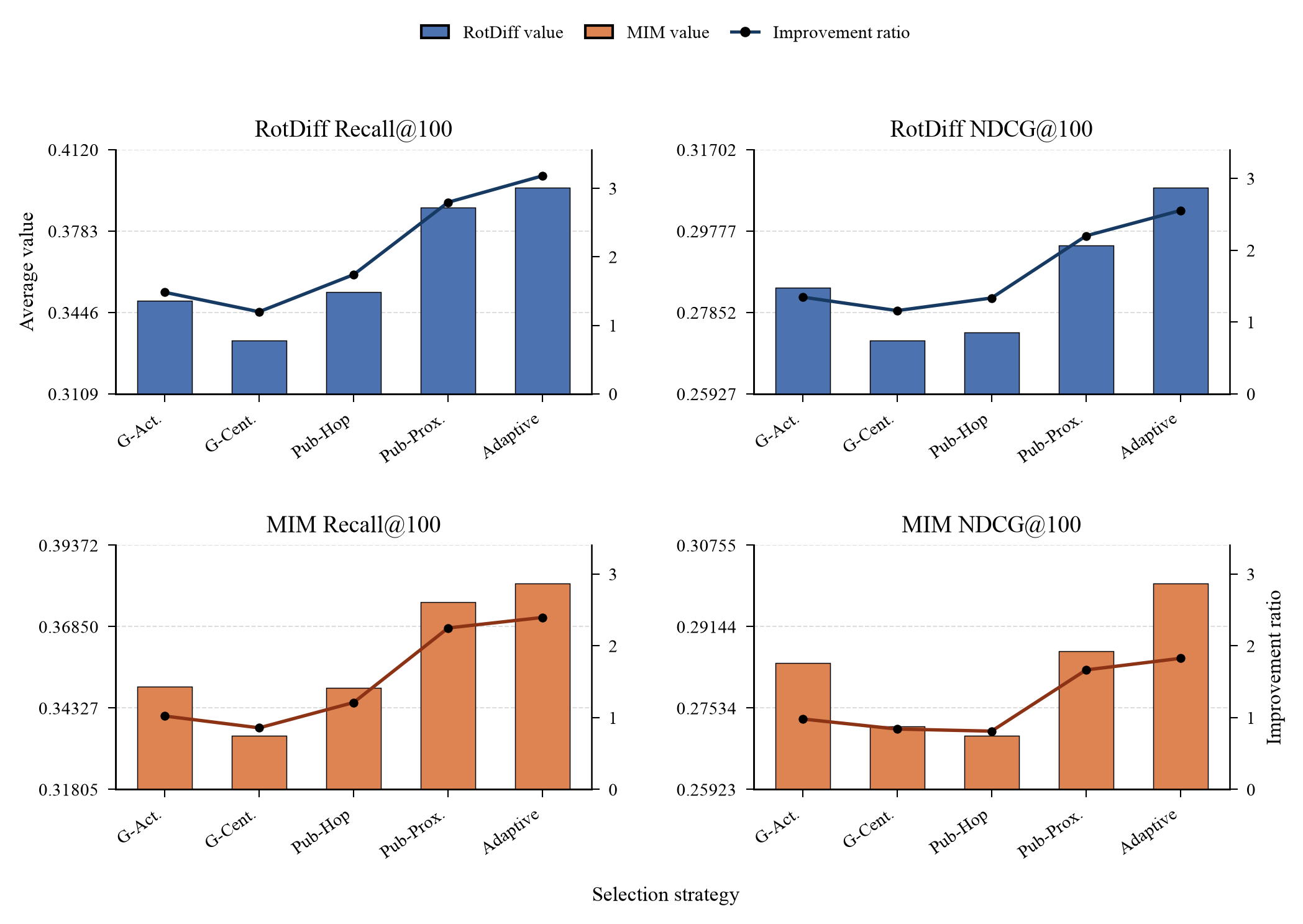}
  \caption{Average Recall and NDCG performance over three datasets, together with the improvement ratios over Random selection strategy of all core-user selection strategies under the RotDiff and MIM backbones at $k{=}100$.}
  \label{fig:selection}
\end{figure}
Figure~\ref{fig:selection} compares the average performance of different core user selection strategies over three datasets under the RotDiff and MIM backbones at $k{=}100$. According to the structure-driven policy, the Adaptive strategy selects Global-Activity for Weibo, Publisher-Hop for Zhihu, and Publisher-Proximity for KuaiRand. Overall, the Adaptive strategy achieves the best results for both Recall and NDCG under both backbones, showing that selecting core users adaptively is more effective than applying a fixed strategy across datasets. The improvement curves further support the advantage of Adaptive selection relative to Random. The result also demonstrates that the benefit of adaptive selection is consistent across two different diffusion backbones, rather than being specific to one structural predictor. 

No single strategy is uniformly optimal across datasets, which motivates the adaptive design of the selection skill. The fixed strategies provide clear gains over Random, but their effectiveness varies with the evaluation metric and backbone. Our dataset-dependent routing reflects the complementary strengths of the candidate strategies: global activity is preferred when overall user activity provides a reliable prioritization signal, publisher-conditioned hop distance is used when direct structural reachability is informative, and publisher proximity is selected when a finer-grained publisher-centered relation is more appropriate. Thus, the Adaptive strategy does not commit to one fixed ranking principle, but chooses among complementary strategies according to the structural characteristics of each dataset.

\subsubsection{Framework Components Ablation}

\begin{table*}[!tbp]
  \centering
  \caption{Component ablation results on three datasets. Each ablation variant is compared with the complete HySID pipeline using the same diffusion backbone. `Sel.' representes the Selection Agent,  `Sim.' represents the Simulation Agent, and `Diff.' represents the Diffusion Agent.}
  \label{tab:ablation-framework-wsdm}
  \small
  \setlength{\tabcolsep}{2.6pt}
  \resizebox{\linewidth}{!}{%
  \begin{tabular}{ll|cc|cc|cc|cc|cc|cc}
    \toprule
    \multirow{3}{*}{Backbone}
      & \multirow{3}{*}{Variant}
      & \multicolumn{4}{c|}{Weibo}
      & \multicolumn{4}{c|}{Zhihu}
      & \multicolumn{4}{c}{KuaiRand} \\
    \cmidrule(lr){3-6}\cmidrule(lr){7-10}\cmidrule(lr){11-14}
      &
      & \multicolumn{2}{c}{Recall} & \multicolumn{2}{c|}{NDCG}
      & \multicolumn{2}{c}{Recall} & \multicolumn{2}{c|}{NDCG}
      & \multicolumn{2}{c}{Recall} & \multicolumn{2}{c}{NDCG} \\
      &
      & @50 & @100 & @50 & @100
      & @50 & @100 & @50 & @100
      & @50 & @100 & @50 & @100 \\
    \midrule

    \multirow{2}{*}{--}
      & w/o Sel. \& Diff.
      & 0.1718 & 0.2168 & 0.1562 & 0.1779
      & 0.0161 & 0.0161 & 0.0134 & 0.0131
      & 0.0107 & 0.0158 & 0.0121 & 0.0124 \\
      & w/o Diff.
      & 0.1671 & 0.1695 & 0.1648 & 0.1662
      & 0.0140 & 0.0140 & 0.0122 & 0.0120
      & 0.0199 & 0.0209 & 0.0197 & 0.0170 \\
    \midrule

    \multirow{3}{*}{RotDiff}
      & w/o Sel. \& Sim.
      & 0.0010 & 0.0150 & 0.0006 & 0.0054
      & 0.0151 & 0.0222 & 0.0069 & 0.0087
      & 0.0053 & 0.0115 & 0.0068 & 0.0088 \\
      & w/o Sim.
      & 0.0988 & 0.1232 & 0.0777 & 0.0869
      & 0.0157 & 0.0221 & 0.0072 & 0.0083
      & 0.0342 & 0.0588 & 0.0331 & 0.0386 \\
      & \cellcolor{gray!15}\textit{HySID}
      & \cellcolor{gray!15}\textbf{0.1793} & \cellcolor{gray!15}\textbf{0.2703*} & \cellcolor{gray!15}\textbf{0.1862} & \cellcolor{gray!15}\textbf{0.2297*}
      & \cellcolor{gray!15}\textbf{0.0353**} & \cellcolor{gray!15}\textbf{0.0468**} & \cellcolor{gray!15}\textbf{0.0236**} & \cellcolor{gray!15}\textbf{0.0272**}
      & \cellcolor{gray!15}\textbf{0.0484} & \cellcolor{gray!15}\textbf{0.0791} & \cellcolor{gray!15}\textbf{0.0411} & \cellcolor{gray!15}\textbf{0.0511} \\
    \midrule

    \multirow{3}{*}{MIM}
      & w/o Sel. \& Sim.
      & 0.0935 & 0.1651 & 0.0703 & 0.1082
      & 0.0285 & 0.0378 & 0.0145 & 0.0154
      & 0.0076 & 0.0125 & 0.0079 & 0.0079 \\
      & w/o Sim.
      & 0.0996 & 0.1274 & 0.0798 & 0.0872
      & 0.0276 & 0.0380 & 0.0137 & 0.0156
      & 0.0308 & 0.0455 & 0.0305 & 0.0324 \\
      & \cellcolor{gray!15}\textit{HySID}
      & \cellcolor{gray!15}\textbf{0.1813*} & \cellcolor{gray!15}\textbf{0.2684*} & \cellcolor{gray!15}\textbf{0.1871} & \cellcolor{gray!15}\textbf{0.2290*}
      & \cellcolor{gray!15}\textbf{0.0356*} & \cellcolor{gray!15}\textbf{0.0481**} & \cellcolor{gray!15}\textbf{0.0233**} & \cellcolor{gray!15}\textbf{0.0270**}
      & \cellcolor{gray!15}\textbf{0.0443} & \cellcolor{gray!15}\textbf{0.0654} & \cellcolor{gray!15}\textbf{0.0372} & \cellcolor{gray!15}\textbf{0.0440} \\
    \bottomrule
  \end{tabular}}
\end{table*}

Table~\ref{tab:ablation-framework-wsdm} compares HySID with four controlled variants. In \textit{w/o Sel. \& Diff.}, the Simulation Agents process all users and the top-$k$ users ranked by simulation score are retained; in \textit{w/o Diff.}, it processes only the selected core users and directly returns their top-$k$ scored users. In \textit{w/o Sel. \& Sim.}, an equally sized set of users is randomly sampled from the full population to form the input cascade, whereas \textit{w/o Sim.} samples the same number of users from the selected core users. Both random-cascade variants retain the top-$k$ diffusion predictions. It worth noting that these two random-cascade variants differ from the standalone diffusion baselines in the overall comparison: to control the cascade size, they provide the diffusion backbone with an input cascade of the same length as the simulation output rather than only the publisher. This design allows us to measure whether the simulated user outcomes improve downstream prediction, instead of attributing the gain merely to a longer input. 

The results first demonstrate the importance of the Diffusion Agent for extending semantic evidence beyond the users explicitly simulated by the LLM. Relative to \textit{w/o Diff.}, the full HySID pipeline achieves higher Recall and NDCG at both cutoffs for both backbones. The improvement is especially pronounced on Zhihu and KuaiRand, where directly ranking the simulated users leaves much of the activated population uncovered. These results indicate that core-user simulation alone can identify informative users, but diffusion is needed to transform this local evidence into a full-population ranking. This result also shows that purely LLM-based approaches do not yield sufficiently accurate individual-level simulations when applied to the full user population. A likely reason is that many users have limited historical interaction records, which provide insufficient behavioral evidence for reliable LLM-based judgment, while LLM-only simulation cannot fully exploit collaborative signals among users.

The comparison between \textit{w/o Sel. \& Sim.} and \textit{w/o Sim.} shows that the location of the input users also matters. Replacing random full-population users with random core users generally improves the diffusion results, especially on Weibo and KuaiRand, indicating that the selection stage provides structurally more suitable candidates even before semantic simulation is introduced. Nevertheless, both random-cascade variants remain below HySID across the reported datasets. Since these variants use input cascades of the same size as the simulation output, it proves that the Simulation Agents contribute informative, user-specific interaction evidence that helps the Diffusion Agent identify and prioritize likely adopters. 

Finally, removing both selection and simulation produces the weakest diffusion-based results in most settings, with particularly large gaps from HySID on Weibo and KuaiRand. Taken together, the ablation results support the complementary role of the three stages: the Selection Agent identifies a structurally informative subset, the Simulation Agents supply item-specific semantic evidence for this subset, and the Diffusion Agent propagates that evidence to generate the final full-population ranking.

\subsection{Parameter Sensitivity Analysis}
\begin{figure}[t]
  \centering
  \includegraphics[width=0.45\textwidth]{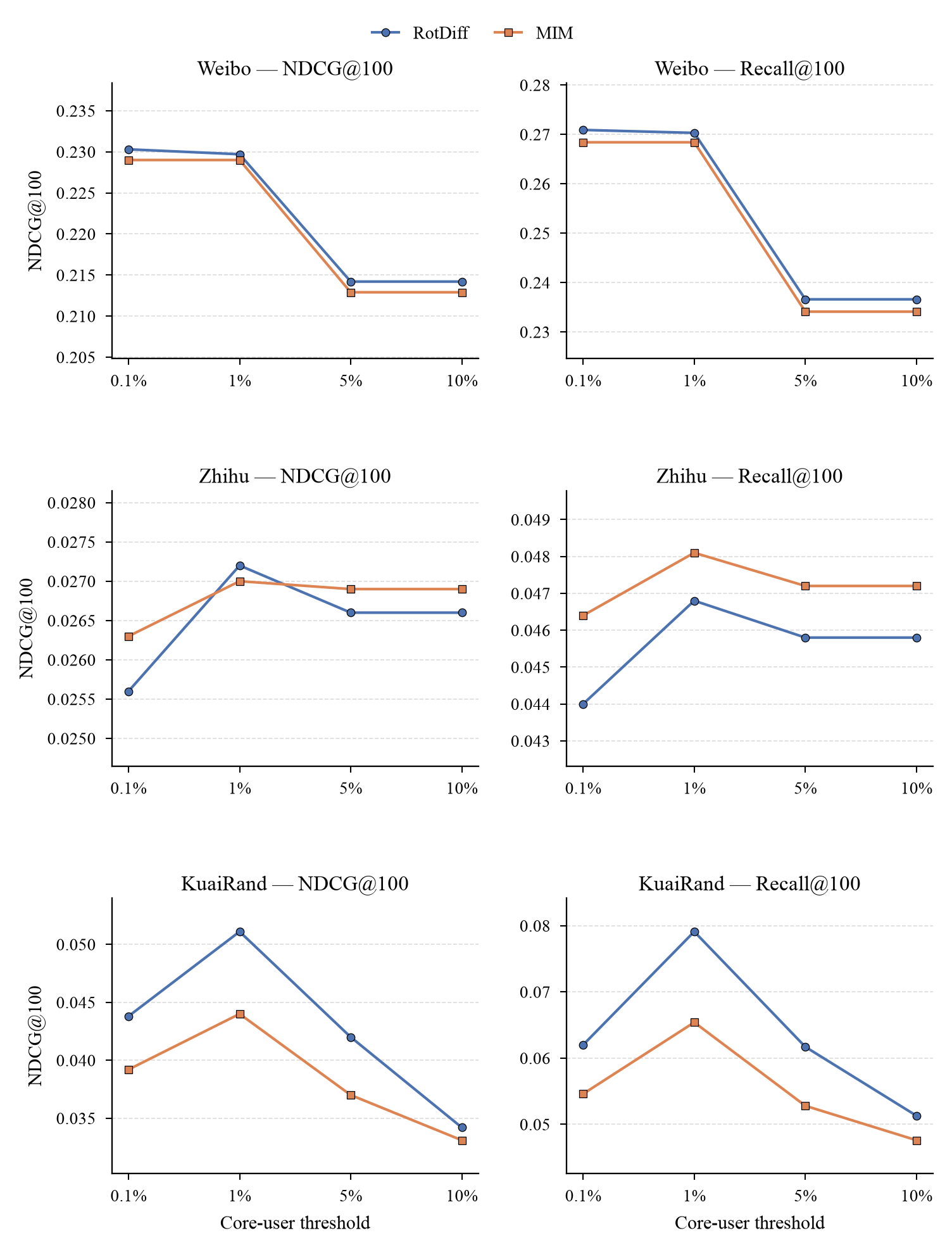}
  \caption{Recall and NDCG sensitivity to the core user selection threshold on three datasets, comparing the RotDiff and MIM backbones at $k=100$.}
  \label{fig:parameter}
\end{figure}
Figure~\ref{fig:parameter} examines the sensitivity of Recall and NDCG to the core user selection threshold under the RotDiff and MIM backbones at $k=100$. The threshold controls the size of the core user set to which the Simulation Agents are applied, and therefore directly affects both the amount of semantic evidence supplied to the Diffusion Agent and the required LLM cost.

Overall, the effect of the threshold is dataset-dependent but consistent across the two backbones. On Weibo, 0.1\% and 1\% yield similar performance, whereas increasing the threshold to 5\% or 10\% reduces both metrics. On Zhihu, performance changes only slightly around 1\%--5\%, suggesting limited benefit from further enlarging the core-user set. KuaiRand exhibits the clearest optimum: both Recall and NDCG peak at 1\% for both backbones and cutoffs, followed by consistent declines at larger thresholds. These trends indicate that selecting more users does not necessarily provide more useful evidence, since an oversized core-user set may introduce less informative simulation results. We therefore use 1\% as a robust default that remains competitive across datasets while limiting LLM computation.

\section{Conclusion}

In this paper, we proposed HySID, a sequential three-agent framework for individual-level information diffusion prediction. The Selection Agent adaptively identifies structure-driven core users, the Simulation Agents generate item-specific semantic judgments for them, and the Diffusion Agent propagates these judgments through a plug-in diffusion backbone to complete the full-population ranking. Experiments on three real-world datasets show that HySID consistently improves multiple diffusion backbones and outperforms general LLM-based simulation baselines, while substantially reducing the LLM cost of full-population simulation. Ablation and strategy analyses further demonstrate the complementary contributions of adaptive selection, semantic simulation, and structural diffusion.



\appendix
\section{Selection skill}
\label{appendix_skill}
Let $n=|\mathcal{U}|$ and let $d^+(u)$ and $d^-(u)$ denote the weighted out- and in-degree of user $u$. The weighted connectivity used by the global centrality tool is computed as the sum of the weighted in- and out-degree. Means, Gini coefficients, component statistics, and related descriptive quantities follow their standard definitions in the literature; the Gini coefficient is applied to both weighted connectivity and historical activity counts.

For publisher $a$, let $C(a)$ denote its weakly connected component. The conditional publisher reachability is \[ r(a)=\frac{|C(a)|-1}{n-1}. \] We report its median over eligible publishers, together with publisher coverage, the isolated-user ratio, the number and size distribution of weakly connected components, and the largest-component ratio. Strongly connected components are computed as a complementary directional-connectivity statistic.

For the comparison between shortest-path and proximity rankings, at most 32 eligible publishers are selected deterministically by sorting their string representations by SHA-256 hash. Let $L_h(a)$ be the candidates at directed hop distance $h$. The hop-tie ratio is \[ \rho_{\mathrm{hop}}(a)=\frac{\sum_h\binom{|L_h(a)|}{2}}{\binom{\sum_h|L_h(a)|}{2}}. \] We report its median when the denominator is nonzero. Personalized PageRank uses publisher personalization, weighted edges, damping $\alpha=0.85$, and the implementation's standard convergence criterion and iteration limit. For each sampled publisher, we compute the Gini coefficient of PPR scores within each non-singleton hop layer and aggregate the layer-level values by layer size. The within-hop PPR dispersion is the median of these aggregated values across the publisher sample. Undefined statistics remain missing and are not imputed. The code for all statistic calculations is available in the released repository.

The registered tools are deterministic implementations of the four strategies defined in the main paper. Invalid or unavailable routing invokes the fixed Global Activity strategy as fallback. The router instruction used by the implementation is shown below. The complete The accompanying \texttt{SKILL.md} is provided in the released code repository. \footnote{\url{https://anonymous.4open.science/r/HySID-4AC6}}

\section{The Construction of behavioral examples used in the Simulation Agent}
\label{appendix_examples}
The positive examples are drawn from observed user interactions. The hard-negative pool is obtained by evaluating a randomly sampled 5\% validation-user subset. And the negative examples are selected up to three items that the target user has not interacted with. The resulting examples and rationales are generated offline before test-time inference. Positive examples are retained only if the assigned score exceeds five; negative examples are retained only if the assigned score falls below five. The exact construction code is released with the repository \footnote{\url{https://anonymous.4open.science/r/HySID-4AC6}}

\section*{Ethical Considerations}
This study models information diffusion from historical interactions on social platforms. All datasets used in this work were de-identified before analysis. The reported experiments use these records only for research-oriented prediction, and do not interpret the results as direct measurements of users' private intentions or psychological states. Individual-level diffusion prediction may nevertheless be misused for surveillance, profiling, or targeted manipulation; HySID is intended for responsible research applications such as understanding diffusion patterns, and its predictions should not support consequential decisions about individuals without appropriate safeguards and human oversight.

Generative AI tools were used only in an assistive role to help implement parts of the authors' manually designed ideas and to edit portions of the manuscript. They were not used to originate the core research ideas, design the method or experimental protocol, or generate the reported results. All AI-assisted code and text were reviewed and revised by the authors, who take full responsibility for the paper.

\bibliographystyle{ACM-Reference-Format}
\bibliography{main}
\end{document}